\documentclass{ws-p8-50x6-00}%
\usepackage{makeidx}
\usepackage{amsmath}
\usepackage{graphicx}%
\usepackage{amsfonts}%
\usepackage{amssymb}

\begin{document}
%
\title{Note on the Path-Integral Variational Approach in the Many-Body Theory}
\author{J. T. Devreese}
\address{Theoretische Fysica van de Vaste Stof,
Universiteit Antwerpen (UIA),
B-2610 Antwerpen, Belgium;
\\also at:
Universiteit Antwerpen (RUCA),
B-2020 Antwerpen, Belgium;
\\ TU Eindhoven,
5600 MB Eindhoven, The Netherlands
\\E-mail: devreese@uia.ua.ac.be}
\maketitle\abstracts
{I discuss how a variatonal approach can be extended to systems of identical
particles (in particular fermions) within the path-integral treatment. The
applicability of the many-body variational principle for path integrals is illustrated for
different model systems, and is shown to crucially depend on whether or not a
model system possesses the proper symmetry with respect to permutations of
identical particles. }%

\section{Introduction}

In the path-integral formulation of quantum mechanics, the so-called
\textit{Jensen-Feynman inequality} provides an upper bound to the free energy of a
quantum system, if properly applied. It was introduced\cite{F55} in Feynman's
path-integral
\index{path-integral} approach to the Fr\"{o}hlich polaron (see formula (8.40)
in Ref.\cite{F72}):
\begin{equation}
F\leq F_{M}+\frac{1}{\beta}\langle S-S_{M}\rangle_{S_{M}}\text{ if }%
S,S_{M}\text{ are real.} \label{JF0}%
\end{equation}

In the variational functional, $F$ and $S$ are the free energy and the action
functional\footnote{It is implicitly assumed that the action functional and
the path integral are expressed in the imaginary-time variable. This
convention is followed throughout the present paper.} of the system under
consideration, whereas $F_{M}$ and $S_{M}$ are the free energy and the action
functional of a model system; the temperature is described by the parameter
$\beta=1/k_{B}T$. Angular brackets mean a weighted average over the
paths\cite{F72}:
\begin{equation}
\langle\bullet\rangle_{S_{M}}\equiv{\frac{\int\bullet\exp(-S_{M}%
)D(\mathrm{path})}{\int\exp(-S_{M})D(\mathrm{path})}}. \label{JF1}%
\end{equation}
A rigorous argument to prove the inequality (\ref{JF0}) is based on the convex
nature of the exponential $\exp(x)$ of a \textit{real} stochastic variable $x$
(see e.g. Fig. 11-1 in Ref. \cite{FH65}), which leads to $\left\langle
e^{x}\right\rangle \geq e^{\left\langle x\right\rangle }$with $\left\langle
x\right\rangle $ the weighted average of $x$.

Apart from Feynman's variational treatment of the ground state energy of a
polaron, the path-integral approach based on the Jensen-Feynman inequality was
successfully applied to a series of problems\cite{SHUL81}, e.g., to the
calculation of the effective classical partition function\cite{FK86,K93} and
of quantum corrections to the free energy of nonlinear systems\cite{GT85,GT86}%
, to the description of all critical exponents observable in second-order
phase transitions\cite{KS-F01}, and to the problem of bipolaron
stability\cite{VPD90,VPD91}.

The derivation of the Jensen-Feynman inequality crucially depends on the
assumption that both the action and the trial action are \textit{real}
functionals. As already recognized by Feynman (see Ref.\cite{FH65}, p. 308), its
application to a \textit{polaron in a magnetic field} therefore becomes
problematic, because the action functional $S$ for a polaron in a magnetic
field (and any reasonable trial action $S_{M}$) is no longer real-valued. A
discussion of this problem lies beyond the scope of the present paper. For
more details on the status of this problem, see the 
literature\cite{HP62,MC70,S76,PD82,L85,DB92,BKP00}.

The Jensen-Feynman inequality is reminiscent of the Bogolubov
inequality\cite{Kvasnikov,D. Ruelle}, which provides the following upper bound
to the free energy $F$ of a system described by the Hamiltonian $H$%
\begin{equation}
F\leq F_{M}-\frac{1}{\beta}\frac{\operatorname*{Tr}\left[  \left(
H-H_{M}\right)  e^{-\beta H_{M}}\right]  }{\operatorname*{Tr}\left(  e^{-\beta
H_{M}}\right)  }\text{ if }H,H_{M}\text{ are Hermitian,} \label{BOGOLUBOV}%
\end{equation}
where $H_{M}$ is the Hamiltonian of some trial system with free energy
$F_{M}.$ The Rayleigh-Ritz variational principle (see e.g. Ref.\cite{S55}, p.
172) for the ground state energy $E\leq\left\langle \Psi_{M}\left|  H\right|
\Psi_{M}\right\rangle /\left\langle \Psi_{M}|\Psi_{M}\right\rangle $ with a
trial state $\left|  \Psi_{M}\right\rangle $ is the zero-temperature limit of
the Bogolubov inequality.

The condition that $S$ and $S_{M}$ are real in (\ref{JF0}) is not necessarily
equivalent to the requirement that $H$ and $H_{M}$ are Hermitian in
(\ref{BOGOLUBOV}). If the Hamiltonians $H$ and $H_{M}$ in the Bogolubov
inequality are Hermitian operators corresponding to Lagrangians $L$ and
$L_{M}$ in the Jensen-Feynman inequality, then the one-to-one correspondence
between (\ref{JF0}) and (\ref{BOGOLUBOV}) guarantees the validity of the
Jensen-Feynman inequality, even if the action functionals are not real (e.g.
for a particle in a magnetic field).

However, both inequalities do not necessarily have the same physical content:
for a system with action $S$ it is not always possible to derive a
corresponding Hamiltonian. For example, for the Fr\"{o}hlich polaron (in the
absence of a magnetic field) the strength of the Jensen-Feynman inequality
lies in the fact that it remains valid after the elimination of the phonons,
with a \textsl{retarded} effective action functional, for which no
corresponding Hamiltonian representation is known. In the operator
formulation, the phonon elimination can formally be realized with
ordered-operator calculus, but this approach involves \textsl{non-Hermitian}
effective operators in the electron variables.

Fermion systems%
\index{fermion systems}%
\index{quantum many-body systems} (with parallel spins) form an important
class of systems for which the Jensen-Feynman inequality is not directly
applicable (whereas the Bogolubov inequality remains valid in the Hilbert
space of antisymmetric states under permutations of the particle coordinates).
The reason is that the path integral for \textsl{fermions} with parallel spin,
\textsl{if expressed in the full coordinate space}, is a superposition of path
integrals with all possible permutations of the particle coordinates, with
negative signs for all odd permutations. For \textsl{bosons}, no negative
signs result from the permutations and the application of the Jensen-Feynman
inequality presents no problems. Therefore, only the many-fermion problem will
be explicitly addressed below.

\section{Path-integral approach for many-body systems}

Recent studies on the path-integral approach to the many-body problem for a
fixed number of \textit{identical particles} by Brosens, Lemmens and
Devreese\cite{LBD96} have allowed to calculate the Feynman-Kac functional on a
state space for $N$ indistinguishable particles, which was found by imposing
an ordering on the configuration space, and the introduction of a set of
boundary conditions in this state space. The path integral
(in the imaginary-time variable) for identical particles was shown to be
positive within this state space. This implies (see subsection \ref{sub:ext}
for more details) that a many-body extension of the Jensen-Feynman inequality
was found, which can be used to evaluate the partition function for
interacting identical particles (Ref.\cite{LBD96}, p. 4476, Ref.\cite{LBD96E}).
This many-body variational principle for path integrals was applied to the
study of thermodynamical properties of a spin-polarized gas of bosons
(Ref.\cite{T98}, abstract, Eq. (3); Ref.\cite{T00}, Eq. (13)). The
applicability of the variational principle as formulated in Ref.\cite{LBD96}
for many-body problems, was discussed in relation to the analysis of
correlations (Ref.\cite{BDL98}, p. 1641) and thermodynamical properties
(Ref.\cite{F99}, p. 3911) of a confined gas of harmonically interacting
spin-polarized fermions.

The many-body variational principle for path integrals%
\index{variational principle} was also used recently in order to calculate the
ground state energy and the optical absorption spectrum of a many-polaron
system, confined to a quantum dot (Ref.\cite{D00}, p. 306).

The remainder of this paper addresses the question \textit{which choice of
model actions is allowed} in order to treat specific systems of interacting
bosons and fermions. I will give some examples, illustrating that the
applicability of the many-body variational principle for path integrals
crucially depends on whether or not a model system possesses the proper
symmetry properties with respect to permutations of identical particles. The
requirements analyzed in this article are qualitatively new as compared
to the Feynman variational principle of Refs.\cite{F55,F72,FH65}.

\section{Many-body variational principle for path integrals}

Let a many-fermion system be described by the action functional
$S[\mathbf{\bar{x}}(t)]$, where $\mathbf{\bar{x}\equiv}\left\{  \mathbf{x}%
_{1},\dots,\mathbf{x}_{N}\right\}  $ are the coordinate vectors of fermions.
The partition function
\index{partition function} $Z_{F}$ of a many-fermion system
\index{many-fermion system} can be expressed as a path integral:
\begin{equation}
Z_{F}=\sum_{P}\frac{\left(  -1\right)  ^{P}}{N!}\int d\mathbf{\bar{x}}%
\int_{\mathbf{\bar{x}}}^{P\mathbf{\bar{x}}}D\mathbf{\bar{x}}\left(  t\right)
\exp\left\{  -S\left[  \mathbf{\bar{x}}\left(  t\right)  \right]  \right\}  ,
\label{Zp}%
\end{equation}
where the summation is over all elements $P$ of the permutation group. The
weight $\left(  -1\right)  ^{P}$ is the character of the representation, i.e.
$+1$ for even permutations and $-1$ for odd permutations (for the case of
fermions). 

\subsection{Model systems with local potentials}

\label{sub:ext} In Ref.\cite{LBD96}, a many-body problem was analyzed for a
local potential $V(\mathbf{\bar{x}})$ (including interparticle interactions)
with the action functional in the imaginary time representation
\begin{equation}
S\left[  \mathbf{\bar{x}}\left(  t\right)  \right]  =\frac{1}{\hbar}\int
_{0}^{\hbar\beta}dt\left[  \frac{m}{2}\sum_{j=1}^{N}\mathbf{\dot{x}}_{j}%
^{2}\left(  t\right)  +V(\mathbf{\bar{x}}\left(  t\right)  )\right]  .
\label{Se0}%
\end{equation}
Note that this action is invariant under the permutations of any two fermions
at any (imaginary) time, since the potential can not make a distinction
between identical particles.

If the potential $V(\mathbf{\bar{x}}\left(  t\right)  )$ is invariant with
respect to the permutations of the \textit{carthesian components of the
particle coordinates}\cite{LBD96,LBD96E}, the many-body propagator was
obtained by four \textit{independent} processes per pair of particles, defined
on a state space ($D_{n}^{3}$ in the notations of Ref.\cite{LBD96}) with
well-defined boundary conditions (see Eqs (4.16--4.17) in Ref.\cite{LBD96} for
details). These processes were shown to give \textit{positive} contributions
to the propagator. Hence, the propagator itself is \textit{positive }on
$D_{n}^{3}$, implying that the Jensen-Feynman inequality can be used to
estimate the partition function for interacting identical
particles\footnote{If the potential is only invariant under permutations of
the particle coordinates, the subprocesses are not linearly independent, and
transitions between the subprocesses have to be taken into account. The actual
analysis in terms of the state space $D_{n}^{3}$ is then only feasible in
practice for a very limited number of fermions. In this case, an overcomplete
space covered by all even permutations of the particle coordinates is more
appropriate.}. The partition function can then be represented 
(apart from a normalizing factor) in the form 
\begin{equation}
Z_{F}=\int_{D_{n}^{3}}d\mathbf{\bar{x}}\int_{\mathbf{\bar{x}}}^{\mathbf{\bar
{x}}}D\mathbf{\bar{x}}\left(  t\right)  \exp\left\{  -S\left[  \mathbf{\bar
{x}}\left(  t\right)  \right]  \right\}  ,\>\>\mathbf{\bar{x}}(t)\in D_{n}%
^{3}. \label{ZpD}%
\end{equation}
Consider now a model system with the action functional in the imaginary time
representation
\begin{equation}
S_{M}\left[  \mathbf{\bar{x}}\left(  t\right)  \right]  =\frac{1}{\hbar}%
\int_{0}^{\hbar\beta}dt\left[  \frac{m}{2}\sum_{j=1}^{N}\mathbf{\dot{x}}%
_{j}^{2}\left(  t\right)  +V_{M}\left(  \mathbf{\bar{x}}\left(  t\right)
\right)  \right]  , \label{Se1}%
\end{equation}
where the \textit{model} potential $V_{M}(\mathbf{\bar{x}})$ contains some
variational parameters, and allows for an analytical calculation of the path
integral. Suppose furthermore that it is invariant with respect to the
permutations of the components of the particles positions. The path-integral
expression for the partition function of the model system is thus:
\begin{equation}
Z_{M}=\int_{D_{n}^{3}}d\mathbf{\bar{x}}\int_{\mathbf{\bar{x}}}^{\mathbf{\bar
{x}}}D\mathbf{\bar{x}}\left(  t\right)  \exp\left\{  -S_{M}\left[
\mathbf{\bar{x}}\left(  t\right)  \right]  \right\}  ,\>\>\mathbf{\bar{x}%
}(t)\in D_{n}^{3}. \label{Z02}%
\end{equation}
One can represent (\ref{ZpD}) as follows:
\begin{align}
Z_{F}  &  \equiv\int_{D_{n}^{3}}d\mathbf{\bar{x}}\int_{\mathbf{\bar{x}}%
}^{\mathbf{\bar{x}}}D\mathbf{\bar{x}}\left(  t\right)  \exp\left\{
-S_{M}\left[  \mathbf{\bar{x}}\left(  t\right)  \right]  -(S\left[
\mathbf{\bar{x}}\left(  t\right)  \right]  -S_{M}\left[  \mathbf{\bar{x}%
}\left(  t\right)  \right]  )\right\}  =\nonumber\\
&  =Z_{M}\langle\exp\left\{  -(S\left[  \mathbf{\bar{x}}\left(  t\right)
\right]  -S_{M}\left[  \mathbf{\bar{x}}\left(  t\right)  \right]  )\right\}
\rangle_{S_{M}}. \label{Zp3}%
\end{align}
Here, the angular brackets denote the quantum statistical expectation value:
\begin{equation}
\langle\bullet\rangle_{S_{M}}\equiv\lbrack Z_{M}]^{-1}\int_{D_{n}^{3}%
}d\mathbf{\bar{x}}\int_{\mathbf{\bar{x}}}^{\mathbf{\bar{x}}}D\mathbf{\bar{x}%
}\left(  t\right)  \bullet\exp\left\{  -S_{M}\left[  \mathbf{\bar{x}}\left(
t\right)  \right]  \right\}  , \label{JF3}%
\end{equation}
analogously to Eq.~\ref{JF1}). The key element of this definition is that the
path integrals in (\ref{Z02}-\ref{JF3}) are defined on the \textit{same} state
space $D_{n}^{3}$, which stems from the symmetry properties of the true action
$S\left[  \mathbf{\bar{x}}\left(  t\right)  \right]  $.

Taking into account that the propagators are positive on the domain $D_{n}%
^{3}$, one obtains the inequality
\begin{equation}
Z_{F}\geq Z_{M}\exp\left\{  -\langle S\left[  \mathbf{\bar{x}}\left(
t\right)  \right]  -S_{M}\left[  \mathbf{\bar{x}}\left(  t\right)  \right]
\rangle_{S_{M}}\right\}  , \label{JF4}%
\end{equation}
which is readily converted into an upper bound for the free energy
\begin{equation}
F_{F}\leq F_{M}+\frac{1}{\beta}\langle S\left[  \mathbf{\bar{x}}\left(
t\right)  \right]  -S_{M}\left[  \mathbf{\bar{x}}\left(  t\right)  \right]
\rangle_{S_{M}}. \label{VAR1}%
\end{equation}
This \textit{many-body variational principle for path integrals}
\index{many-body inequality for path integrals} is formally very similar to
the Jensen-Feynman inequality (\ref{JF0}). The difference between Eqs.~(\ref{VAR1})
and (\ref{JF0}) lies in the definition of the expectation values. In
(\ref{JF4}) and (\ref{VAR1}) the expectation value (\ref{JF3}) is defined over
a subdomain $D_{n}^{3}$ of the configuration space, whereas the expectation
value (\ref{JF1}) in (\ref{JF0}) is defined over the full configuration space.
However, because the symmetry properties allow to unfold the state space into
the full configuration space, the restriction to the state space can be
omitted in the calculation. The state space ($D_{n}^{3}$ in this example) only
serves the goal to check whether the action and the trial action have the
correct symmetry properties.

\subsection{Model systems with retarded effective interactions}

\label{sub:ret} The action functional $S[\mathbf{\bar{x}}(t)]$ of the system
under study can contain a retarded effective interaction.
\index{retarded interaction} This is the case, e.g., for a system of $N$
polarons after the phonon variables have been integrated out. Such
many-fermion systems substantially differ from those considered above in
subsection \ref{sub:ext}, and a different class of model systems seems
appropriate. For this purpose, we consider a model system consisting of
fermions in interaction with \textit{auxiliary fictitious particles}.

Such a model system has the action functional in the imaginary-time
representation
\begin{equation}
S_{M}=\frac{1}{\hbar}\int_{0}^{\hbar\beta}L_{M}\left(  t\right)  dt.
\label{SM}%
\end{equation}
The model ``Lagrangian'' is chosen in the form:
\begin{equation}
L_{M}=L_{F}(\mathbf{\bar{x}})+L_{f}(\mathbf{\bar{y}})+L_{F-f}(\mathbf{\bar{x}%
},\mathbf{\bar{y}}), \label{LM}%
\end{equation}
where $\left\{  \mathbf{x}_{j}\right\}  \equiv\mathbf{\bar{x}}$ are the
coordinate vectors of the fermions, and $\left\{  \mathbf{y}_{j}\right\}
\equiv\mathbf{\bar{y}}$ are the coordinate vectors of the fictitious
particles. The ``Lagrangians'' $L_{F}(\mathbf{\bar{x}})$, $L_{f}%
(\mathbf{\bar{y}})$ and $L_{F-f}(\mathbf{\bar{x}},\mathbf{\bar{y}})$ describe fermions, fictitious particles and the interaction between the
fermions and the fictitious particles,
respectively. Here, the
discussion is limited to the case of \textit{distinguishable} fictitious
particles for the sake of simplicity.

The partition function $Z_{M}$ of the model system can be written as the
following path integral:
\begin{equation}
Z_{M}=\sum_{P}\frac{\xi^{P}}{N!}\int d\mathbf{\bar{x}}\int d\mathbf{\bar{y}%
}\int_{\mathbf{\bar{x}}}^{P\mathbf{\bar{x}}}D\mathbf{\bar{x}}\left(  t\right)
\int_{\mathbf{\bar{y}}}^{\mathbf{\bar{y}}}D\mathbf{\bar{y}}\left(  t\right)
\exp\left\{  -S_{M}\right\}  . \label{Z01A}%
\end{equation}
Integrating out the coordinates of the fictitious particles, the partition
function (\ref{Z01A}) takes the form
\begin{align}
Z_{M}  &  =Z_{0}Z_{f},\label{ZM}\\
Z_{0}  &  =\sum_{P}\frac{\xi^{P}}{N!}\int d\mathbf{\bar{x}}\int_{\mathbf{\bar
{x}}}^{P\mathbf{\bar{x}}}D\mathbf{\bar{x}}\left(  t\right)  \exp\left\{
-S_{0}\left[  \mathbf{\bar{x}}\left(  t\right)  \right]  \right\}  ,
\label{ZpM}%
\end{align}
where $Z_{f}$ is the partition function of the system of fictitious
particles:
\begin{equation}
Z_{f}=\int d\mathbf{\bar{y}}\int_{\mathbf{\bar{y}}}^{\mathbf{\bar{y}}%
}D\mathbf{\bar{y}}\left(  t\right)  \exp\left\{  -\frac{1}{\hbar}\int
_{0}^{\hbar\beta}L_{f}(\mathbf{\bar{y}})\right\}  , \label{Z02A}%
\end{equation}
and $S_{0}\left[  \mathbf{\bar{x}}\left(  t\right)  \right]  $ is an effective
action which only depends on the fermion variables%
\begin{equation}
S_{0}\left[  \mathbf{\bar{x}}\left(  t\right)  \right]  \equiv\frac{1}{\hbar
}\int_{0}^{\hbar\beta}\left(  L_{F}\left(  t\right)  dt+\Phi_{0}\left[
\mathbf{\bar{x}}\left(  t\right)  \right]  \right)  . \label{IPD}%
\end{equation}
The last term in (\ref{IPD}) is referred to as the \textit{influence
phase} of the fictitious particles. It is defined as
\begin{align}
&  \exp\left\{  -\Phi_{0}\left[  \mathbf{\bar{x}}\left(  t\right)  \right]
\right\}  \equiv\nonumber\\
&  \equiv\lbrack Z_{f}]^{-1}\int d\mathbf{\bar{y}}\int_{\mathbf{\bar{y}}%
}^{\mathbf{\bar{y}}}D\mathbf{\bar{y}}\left(  t\right)  \exp\left\{
-\frac{1}{\hbar}\int_{0}^{\hbar\beta}\left[  L_{f}(\mathbf{\bar{y}}%
)+L_{e-f}(\mathbf{\bar{x}},\mathbf{\bar{y}})\right]  dt\right\}.  \label{FAZA}%
\end{align}
For interaction Lagrangians $L_{F-f}(\mathbf{\bar{x}},\mathbf{\bar{y}})$ which
are quadratic in $\mathbf{\bar{y}}$, the influence phase can be shown to take
the form of a \textit{retarded effective interaction}:
\begin{equation}
\Phi_{0}\left[  \mathbf{\bar{x}}(t)\right]  =\int\limits_{0}^{\hbar\beta
}dt\int\limits_{0}^{\hbar\beta}dsK(t,s)\mathbf{X}\left(  t\right)
\cdot\mathbf{X}\left(  s\right)  , \label{FAZA2}%
\end{equation}
where $K(t,s)$ depends on two time variables $(t,s)$, while $\mathbf{X}\left(
t\right)  $ is a linear function of the fermion coordinates $\mathbf{\bar{x}%
}\left(  t\right)  $ (see the next section for specific examples).

If the action functional $S\left[  \mathbf{\bar{x}}\left(  t\right)  \right]
$ of the system under study satisfies the permutation symmetry conditions
discussed in the previous subsection, its partition function can
be represented as a path integral over the space state $D_{n}^{3}$ -- in the
form (\ref{ZpD}):
\begin{align}
Z_{F}  &  \equiv\int_{D_{n}^{3}}d\mathbf{\bar{x}}\int_{\mathbf{\bar{x}}%
}^{\mathbf{\bar{x}}}D\mathbf{\bar{x}}\left(  t\right)  \exp\left\{
-S_{0}\left[  \mathbf{\bar{x}}\left(  t\right)  \right]  -(S\left[
\mathbf{\bar{x}}\left(  t\right)  \right]  -S_{0}\left[  \mathbf{\bar{x}%
}\left(  t\right)  \right]  )\right\}  ,\nonumber\\
&  =Z_{0}\langle\exp\left\{  -(S\left[  \mathbf{\bar{x}}\left(  t\right)
\right]  -S_{0}\left[  \mathbf{\bar{x}}\left(  t\right)  \right]  )\right\}
\rangle_{S_{0}}. \label{Zp2}%
\end{align}
If the model action $S_{0}\left[  \mathbf{\bar{x}}\left(  t\right)  \right]  $
also possesses the above symmetry properties with respect to permutations, its
partition function (\ref{ZpM}) can also be written in the form of a path
integral over the domain $D_{n}^{3}$,
\begin{equation}
Z_{0}=\int_{D_{n}^{3}}d\mathbf{\bar{x}}\int_{\mathbf{\bar{x}}}^{\mathbf{\bar
{x}}}D\mathbf{\bar{x}}\left(  t\right)  \exp\left\{  -S_{0}\left[
\mathbf{\bar{x}}\left(  t\right)  \right]  \right\}  , \label{ZpMD}%
\end{equation}
and a quantum statistical expectation value can be defined in (\ref{Zp2})
\begin{equation}
\langle\bullet\rangle_{S_{0}}\equiv\lbrack Z_{0}]^{-1}\int_{D_{n}^{3}%
}d\mathbf{\bar{x}}\int_{\mathbf{\bar{x}}}^{\mathbf{\bar{x}}}D\mathbf{\bar{x}%
}\left(  t\right)  \bullet\exp\left\{  -S_{0}\left[  \mathbf{\bar{x}}\left(
t\right)  \right]  \right\}  , \label{JF2}%
\end{equation}
analogously to the definition (\ref{JF1}). The fact that the propagators are
positive in the integration domain $D_{n}^{3}$ guarantees that the inequality%
\begin{equation}
Z_{F}\geq Z_{0}\exp\left\{  \langle-(S\left[  \mathbf{\bar{x}}\left(
t\right)  \right]  -S_{0}\left[  \mathbf{\bar{x}}\left(  t\right)  \right]
)\rangle_{S_{0}}\right\}  \label{JF4l}%
\end{equation}
holds true. Consequently we obtain an upper bound for the free energy, similar
to Eq.~(\ref{VAR1}):%
\begin{equation}
F_{F}\leq F_{v}\equiv F_{0}+{\frac{1}{\beta}}\langle S\left[  \mathbf{\bar{x}%
}\left(  t\right)  \right]  -S_{0}\left[  \mathbf{\bar{x}}\left(  t\right)
\right]  \rangle_{S_{0}}. \label{VAR}%
\end{equation}
where $F_{0}=-{\frac{1}{\beta}}\ln Z_{0}$ and $F_{F}=-{\frac{1}{\beta}}\ln
Z_{F}$.

Because of the presence of the retarded action (\ref{FAZA2}) resulting from
the elimination of the fictitious particles, one should guarantee that the
functional $S_{0}\left[  \mathbf{\bar{x}}\left(  t\right)  \right]  $ has the
required symmetry with respect to permutations which allows that the many-body
processes, related to the quantum statistical expectation value (\ref{JF2}),
are restricted to the state space $D_{n}^{3}$ at any time. This condition is
an essential ingredient for the justification of the \textit{many-body
variational principle for path integrals} (\ref{VAR1}). Like in the case
(\ref{VAR1}) of local potentials, the upper bound (\ref{VAR}) to the free
energy is formally very similar to the Jensen-Feynman inequality (\ref{JF0}):
the difference lies again in the definition of the expectation values. In
(\ref{JF4l}) and (\ref{VAR}) the expectation value (\ref{JF2}) is defined over
the subdomain $D_{n}^{3}$ of the configuration space, whereas the expectation
value (\ref{JF1}) in (\ref{JF0}) is defined over the full configuration space.
However, like in the previous subsection, the state space ($D_{n}^{3}$ in this
case) is only needed to check whether the symmetry of the action and the trial
action allows to apply the inequality. The calculation can be performed over
the total configuration space by unfolding the state space.

\section{Examples: non-interacting fermions as a test case for a many-body
variational principle with path integrals}

\subsection{Model system, in which each fermion harmonically interacts with
one fictitious particle}

\label{sub:1}In order to illustrate the applicability of the many-body
variational principle for path integrals (\ref{VAR}), and in particular the
need of the correct symmetry requirements for the model action, we first
consider a very simple system of $N=\sum_{\sigma=\pm1/2}N_{\sigma}$
non-interacting fermions, described by the Lagrangian%
\begin{equation}
L_{F}=\frac{m}{2}\sum_{\sigma=\pm1/2}\sum_{j=1}^{N_{\sigma}}\mathbf{\dot{x}%
}_{j,\sigma}^{2}, \label{L1}%
\end{equation}
where $N_{\sigma}$ is the number of electrons with spin component $\sigma
=\pm1/2$. The ground state energy for the system with classical ``Lagrangian''
(\ref{L1}) is elementary:
\begin{equation}
E_{0}=\frac{3}{5}E_{F}\quad\left(  E_{F}\equiv\frac{\hbar^{2}k_{F}^{2}}%
{2m}\right)  \label{FREE}%
\end{equation}
with the Fermi wave number $k_{F}$ and the Fermi energy $E_{F}$.

We now examine whether the many-body variational principle for path integrals
(\ref{VAR}) indeed provides an upper bound to the correct ground state energy
(\ref{FREE}).

For the model system (\ref{LM}) we choose a Lagrangian $L_{M}$, in which each
fermion harmonically interacts with \textit{one fictitious particle.} We do
this uncritically, deliberately overlooking the problem of the required
symmetry of the state space of the model action and choose
\begin{align}
L_{M}  &  =\frac{m}{2}\sum_{\sigma=\pm1/2}\sum_{j=1}^{N_{\sigma}}%
\mathbf{\dot{x}}_{j,\sigma}^{2}+\nonumber\\
&  +\frac{M}{2}\sum_{\sigma=\pm1/2}\sum_{j=1}^{N_{\sigma}}\mathbf{\dot{y}%
}_{j,\sigma}^{2}+\frac{k}{2}\sum_{\sigma=\pm1/2}\sum_{j=1}^{N_{\sigma}}\left(
\mathbf{x}_{j,\sigma}-\mathbf{y}_{j,\sigma}\right)  ^{2}, \label{LMone}%
\end{align}
where $M$ is the mass of the fictitious particles, and $k$ is the force
constant of the elastic bond between a fermion and its accompanying fictitious
particle. We introduce the following notations:
\[
w=\sqrt{\frac{k}{M}},\>\>\>v=\sqrt{\frac{k}{\mu}},\>\>\>\mu=\frac{mM}{m+M}.
\]
For this particular case, the elimination of the fictitious particles leads to
the influence phase (\ref{FAZA}):
\begin{equation}
\Phi_{0}\left[  \mathbf{\bar{x}}(t)\right]  =-\frac{Mw^{3}}{8\hbar}%
\int\limits_{0}^{\hbar\beta}dt\int\limits_{0}^{\hbar\beta}ds\frac{\cosh
w\left(  \left|  t-s\right|  -\frac{\hbar\beta}{2}\right)  }{\sinh\frac{1}%
{2}\hbar\beta w}\sum_{\sigma=\pm1}\sum_{j=1}^{N_{\sigma}}[\mathbf{x}%
_{j,\sigma}\left(  t\right)  -\mathbf{x}_{j,\sigma}\left(  s\right)  ]^{2},
\label{Fi0}%
\end{equation}
with a quadratic effective retarded self interaction for each fermion.

\begin{figure}[h]
\epsfxsize     =20pc
\centerline{
\epsfbox{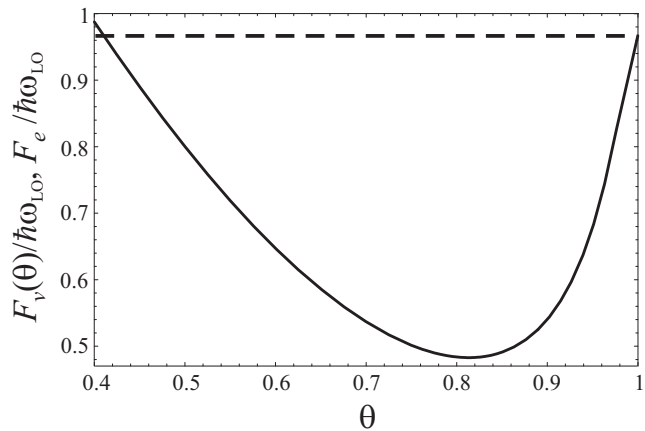} }  \caption{Free energy $F_{v}(\theta)$ ($\theta=w/v$) for
non-interacting fermions [according to the inequality (\ref{VAR})]
 (solid line) compared with the exact free
energy $F_{\mathrm{{e}}}$ (dashed line) at the optimal value of the
variational frequency parameter $v=1.83$ (in units of $E_{\mathrm{{F}}}/\hbar
$, where $E_{\mathrm{{F}}}=49$ meV is the Fermi energy for the density
$n_{e}=5\times10^{19}$ cm$^{-3},$ rather arbitrarily chosen); $\beta=16.4$ (in
units of $E_{\mathrm{{F}}}^{-1}$). The mass of a fermion is taken $m=m_{0},$
the bare electron mass.}%
\label{fig:1}%
\end{figure}

Then we apply the many-body variational principle for path integrals
(\ref{VAR}) naively for the chosen model system. The free energy $F_{v}$
from this inequality is calculated analytically. The parameters $M$ and $k$ of
the model ``Lagrangian'' (\ref{LM}) are then found by minimizing the value of
the supposed upper bound $F_{v}$. This calculation has shown that the
many-body variational principle for path integrals (\ref{VAR}) is
\textit{violated} for this model system, as clearly illustrated in
Fig.~\ref{fig:1}.

What was wrong in the above approach? The answer is immediate: ``Of course, we
forgot to check whether the model action has the required symmetry!'' The
model ``Lagrangian'' (\ref{LMone}) is \textit{not symmetrical} with respect to
the permutations of the fermion coordinates $\mathbf{x}_{j,\sigma}$, because
each of them is linked with a particular fictitious particle.

\subsection{Model system, in which each fermion has equal elastic bonds with
all fictitious particles}

\label{sub:3}

As a second example, we study the many-body variational principle for path
integrals (\ref{VAR}) for $N=\sum_{\sigma=\pm1/2}N_{\sigma}$ non-interacting
electrons in a parabolic confinement, described by the Lagrangian (in the
imaginary-time variable)%
\begin{equation}
L=\frac{m}{2}\sum_{\sigma=\pm1/2}\sum_{j=1}^{N_{\sigma}}\left(  \mathbf{\dot
{x}}_{j,\sigma}^{2}+\Omega_{0}^{2}\mathbf{x}_{j,\sigma}^{2}\right)  .
\label{L}%
\end{equation}
In particular, the case $\Omega_{0}\rightarrow0$ will be analyzed (a
translationally invariant system, equivalent to the free-fermion model in
subsection \ref{sub:1}). The exact ground-state energy of this system can
immediately be written down:
\begin{equation}
E_{0}\left(  \Omega_{0},N\right)  =\hbar\Omega_{0}\sum_{\sigma=\pm1/2}\left\{
N_{\sigma}\left[  n_{0}\left(  N_{\sigma}\right)  +\frac{5}{2}\right]
-\frac{1}{4}N_{0}\left(  n_{0}\left(  N_{\sigma}\right)  \right)  \left[
n_{0}\left(  N_{\sigma}\right)  +4\right]  \right\}  . \label{E0ex}%
\end{equation}
Here, $n_{0}(N_{\sigma})$ is the number of the upper fully occupied energy
level for $N_{\sigma}$ fermions with spin component $\sigma$. The number of
fermions in all closed shells [$N_{0}(n_{0}(N_{\sigma}))\leq N_{\sigma}$] is
\begin{equation}
N_{0}\left(  n_{0}\right)  \equiv\sum_{n=0}^{n_{0}}\frac{\left(  n+1\right)
\left(  n+2\right)  }{2}=\frac{1}{6}\left(  n_{0}+1\right)  \left(
n_{0}+2\right)  \left(  n_{0}+3\right)  . \label{4}%
\end{equation}

A model system is now considered, which consists of particles in a harmonic
confinement potential with elastic interparticle interactions as studied in
Ref.\cite{BDL97}. The ``Lagrangian'' of this model system is chosen in the
form
\begin{align}
L_{M}  &  =\frac{m}{2}\sum_{\sigma=\pm1/2}\sum_{j=1}^{N_{\sigma}}\left(
\mathbf{\dot{x}}_{j,\sigma}^{2}+\Omega^{2}\mathbf{x}_{j,\sigma}^{2}\right)
+\frac{m\omega^{2}}{4}\sum_{\sigma=\pm1/2}\sum_{j=1}^{N_{\sigma}}\sum
_{\tau=\pm1/2}\sum_{l=1}^{N_{\sigma}}\left(  \mathbf{x}_{j,\sigma}%
-\mathbf{x}_{l,\tau}\right)  ^{2}\nonumber\\
&  +\frac{M}{2}\sum_{l=1}^{N_{B}}\left(  \mathbf{\dot{y}}_{j}^{2}+\Omega
_{B}^{2}\mathbf{y}_{j}^{2}\right)  +\frac{M\omega_{B}^{2}}{4}\sum_{j=1}%
^{N_{B}}\sum_{l=1}^{N_{B}}\left(  \mathbf{y}_{j}-\mathbf{y}_{l}\right)
^{2}\nonumber\\
&  +\frac{k}{2}\sum_{\sigma=\pm1/2}\sum_{j=1}^{N_{\sigma}}\sum_{l=1}^{N_{B}%
}\left(  \mathbf{x}_{j,\sigma}-\mathbf{y}_{l}\right)  ^{2}. \label{LAG}%
\end{align}
The frequencies $\Omega,$ $\omega,$ $\Omega_{B},$ $\omega_{B},$ the mass $M$
of a fictitious particles$,$ and the force constant $k$ are treated as
\textit{variational parameters}. It is important to stress, that in this model
system each fermion has identical elastic bonds with\textit{ all }fictitious
particles, and therefore permutations of the fermion coordinates leave the
``Lagrangian'' (\ref{LAG}) invariant.

After integration over the paths of the fictitious particles, the partition
function (\ref{ZM}) becomes
\begin{align}
Z_{f}  &  =\left[  \sinh\left(  {\frac{\hbar\beta\tilde{\Omega}_{B}}{2}%
}\right)  \right]  ^{3}\left[  \sinh\left(  {\frac{\hbar\beta w_{B}}{2}%
}\right)  \right]  ^{3N_{B}-3},\\
\text{with }\tilde{\Omega}_{B}  &  =\sqrt{\Omega^{2}+\frac{kN_{B}}{m}%
},\>\>w_{B}=\sqrt{\Omega^{2}-N_{B}\omega^{2}+\frac{kN_{B}}{m}}. \label{OM3}%
\end{align}
The action (\ref{IPD}) in the imaginary time representation takes the form
\begin{align}
S_{0}\left[  \mathbf{\bar{x}}\left(  t\right)  \right]   &  \equiv
S_{F0}\left[  \mathbf{\bar{x}}\left(  t\right)  \right]  +\Phi_{0}\left[
\mathbf{\bar{x}}\left(  t\right)  \right]  ,\label{S00}\\
S_{F0}\left[  \mathbf{\bar{x}}\left(  t\right)  \right]   &  =\frac{1}{\hbar
}\int_{0}^{\hbar\beta}dt\left[  \sum_{\sigma=\pm1/2}\sum_{j=1}^{N_{\sigma}%
}\frac{m}{2}\mathbf{\dot{x}}_{j,\sigma}^{2}\left(  t\right)  +\frac{m\omega
^{2}N^{2}}{2}\mathbf{X}^{2}\right]  ,\\
\text{with }\mathbf{X}  &  \equiv\sum_{\sigma=\pm1/2}\sum_{j=1}^{N_{\sigma}%
}\mathbf{x}_{j,\sigma}, \label{Se3}%
\end{align}
and the ``influence phase'' (\ref{FAZA}) for the present model becomes:
\begin{equation}
\Phi_{0}\left[  \mathbf{\bar{x}}(t)\right]  =\frac{k^{2}N^{2}N_{B}^{2}}%
{4m_{B}\hbar\tilde{\Omega}_{B}}\int\limits_{0}^{\hbar\beta}dt\int
\limits_{0}^{\hbar\beta}ds\frac{\cosh\left[  \tilde{\Omega}_{B}\left(  \left|
t-s\right|  -{\frac{\hbar\beta}{2}}\right)  \right]  }{\sinh\left(
{\frac{\hbar\beta\tilde{\Omega}_{B}}{2}}\right)  }\mathbf{X}\left(  t\right)
\cdot\mathbf{X}\left(  s\right)  . \label{Fi03}%
\end{equation}
As distinct from (\ref{LMone}), this model Lagrangian (\ref{LAG}) is
\textit{invariant with respect to permutations of the components of all
fermion coordinates}. Hence, for the chosen model system the symmetry
conditions on the action are fulfilled (as formulated in subsection
\ref{sub:1}), which ensures the validity of the many-body variational
principle for path integrals (\ref{VAR}).

The functional [resulting from (\ref{VAR}) at zero temperature] for the ground
state energy of $N$ fermions in a parabolic confinement with the confinement
frequency $\Omega_{0}$ takes the form
\begin{align}
&  E_{v}\left(  \Omega_{1},\Omega_{2},w,w_{B}\right)  =\hbar\left\{
\frac{\Omega_{0}^{2}+w^{2}}{2w^{2}}\left[  \tilde{E}\left(  w,N\right)
-\frac{3}{2}w\right]  +\frac{3}{2}\left(  \Omega_{1}+\Omega_{2}-w_{B}\right)
\right. \nonumber\\
&  +\left.  \frac{3}{4}\left(  \Omega_{0}^{2}-\Omega_{1}^{2}-\Omega_{2}%
^{2}+w_{B}^{2}\right)  \sum_{i=1}^{2}\frac{a_{i}^{2}}{\Omega_{i}%
}+\frac{3\gamma^{2}}{4w_{B}}\sum_{i=1}^{2}\frac{a_{i}^{2}}{\Omega_{i}\left(
\Omega_{i}+w_{B}\right)  }\right\}  , \label{ENER}%
\end{align}
where the following notations are used:
\begin{align}
&  a_{1}=\left(  \frac{\Omega_{1}^{2}-w_{B}^{2}}{\Omega_{1}^{2}-\Omega_{2}%
^{2}}\right)  ^{1/2},\quad a_{2}=\left(  \frac{w_{B}^{2}-\Omega_{2}^{2}%
}{\Omega_{1}^{2}-\Omega_{2}^{2}}\right)  ^{1/2},\\
&  \gamma=\left[  \left(  \Omega_{1}^{2}-w_{B}^{2}\right)  \left(  w_{B}%
^{2}-\Omega_{2}^{2}\right)  \right]  ^{1/2},\\
&
\begin{array}
[b]{l}%
\tilde{E}\left(  w,N\right)  =\\
=\hbar w\sum\limits_{\sigma=\pm1/2}\left\{  N_{\sigma}\left[  n_{0}\left(
N_{\sigma}\right)  +\frac{5}{2}\right]  -\frac{1}{4}N_{0}\left(  n_{0}\left(
N_{\sigma}\right)  \right)  \left[  n_{0}\left(  N_{\sigma}\right)  +4\right]
\right\}  .
\end{array}
\end{align}

The difference between the upper bound to the ground-state energy (\ref{ENER})
and the exact ground-state energy (\ref{E0ex}) is clearly positive:
\begin{equation}
E_{v}\left(  \Omega_{1},\Omega_{2},w,w_{B}\right)  -E_{0}\left(  \Omega
_{0},N\right)  =\frac{3\hbar}{4}\frac{\left(  \Omega_{0}-z\right)  ^{2}}%
{z}\label{ac} 
\end{equation}
with
\begin{equation}
z=\frac{\Omega_{1}\Omega_{2}+\Omega_{0}^{2}}{\Omega
_{1}+\Omega_{2}}, \nonumber
\end{equation}
and it follows from (\ref{ac}) that the inequality
\begin{equation}
E_{v}\left(  \Omega_{1},\Omega_{2},w,w_{B}\right)  \geq E_{0}\left(
\Omega_{0},N\right)  \label{ENER2}%
\end{equation}
holds true for any values of the variational parameters, \textit{in accordance
with the many-body variational principle for path integrals} (\ref{VAR}).

The minimal value of the functional (\ref{ENER}), which is achieved at
$z=\Omega_{0}$, coincides with the exact ground-state energy (\ref{E0ex}) of
$N$ non-interacting fermions in a parabolic confinement with the frequency
$\Omega_{0}$. This result confirms the applicability of the many-body
variational principle for path integrals (\ref{VAR}) to the many-fermion
system under consideration, with a model system, whose ``Lagrangian''
(\ref{LAG}) is symmetric under the permutations of the components of the
fermion positions.

Thus, the many-body variational principle for path integrals (\ref{VAR}) is
satisfied for a system of non-interacting fermions in a parabolic confinement
potential (including the translationally invariant case $\Omega_{0}=0$), when
the model system with the ``Lagrangian'' (\ref{LAG}) is considered.

Clearly, the Lagrangian (\ref{L}) was not considered for its own sake, since
it is trivial to treat. It was merely presented as a test case to illustrate
our many-body variational principle for path integrals with two non-trivial
trial actions which elucidate the crucial role of the correct symmetry requirements.

\section{Conclusions}

Summarizing, the applicability of the many-body variational principle for path
integrals (\ref{VAR}) crucially depends on the symmetry of the model system
with respect to permutations of identical particles.

The invariance of the action functional $S_{0}[\mathbf{\bar{x}}(t)]$ and of
the model action with respect to permutations of components of the positions
of any two fermions at any (imaginary) time ensures that the fermion
propagator is positive on the state space $D_{n}^{3}$.

It should be noted that these rather stringent symmetry conditions are used as
an example. A more general variational principle for identical particles, not
limited to $D_{n}^{3}$ but to a much larger subspace of the configuration
space, will be presented in future publications.

The main result of the present analysis is that a many-body variational
principle for path integrals (\ref{VAR}) can be found in the framework of the
many-body path integral approach, even for retarded effective interactions,
provided that the model action and the true action have the appropriate
symmetry properties under permutations of the particle coordinates.

\section*{Acknowledgments}

I like to thank V. M. Fomin for interesting discussions during the preparation
of the manuscript, and S. N. Klimin for assistance with the numerical aspects.
Stimulating interaction with F.~Brosens and L.~Lemmens in the frame of our
collaborations on path integrals for many-body systems is gratefully
acknowledged. This work has been supported by the BOF NOI (UA-UIA), GOA BOF UA
2000, IUAP, FWO-V projects G.0287.95, G.0071.98, G.0274.01N and the W.O.G.
WO.025.99N (Belgium).

\end{document}